\documentclass{conm-p-l}

\usepackage{graphicx}
\usepackage{latexsym}
\usepackage{longtable}

\newcommand{\Rn}{{\mathbb{R}}}

\newcommand{\ft}[2]{{\textstyle {\frac{#1}{#2}} }}

\begin{document}

\title[Low Level Representations for ${E_{10}}$ and ${E_{11}}$]{Low Level Representations for ${E_{10}}$ and ${E_{11}}$.} 
\author[H. Nicolai]{Hermann Nicolai}
\address{Max-Planck-Institut f\"ur Gravitationsphysik, 
Albert Einstein Institut, M\"uhlenberg 1, D-14476 Potsdam, Germany}
\email{nicolai@aei.mpg.de}
\author[T. Fischbacher]{Thomas Fischbacher}
\email{tf@aei.mpg.de}

\begin{abstract}
We work out the decomposition of the indefinite Kac Moody algebras
${E_{10}}$ and ${E_{11}}$ w.r.t. their respective subalgebras $A_9$ and $A_{10}$
at low levels. Tables of the irreducible representations with their 
outer multiplicities are presented for ${E_{10}}$ up to level $\ell = 18$ 
and for ${E_{11}}$ up to level $\ell =10$. On the way we confirm and extend 
existing results for ${E_{10}}$ root multiplicities, and for the first time 
compute non-trivial root multiplicities of ${E_{11}}$.
\end{abstract}

\maketitle

\section{Introduction}

In this contribution, we will be mainly concerned with the indefinite
Kac Moody algebras $E_{10}$ and $E_{11}$ (see \cite{Kac,MP} for the basic 
theory of Kac Moody algebras). Of these $E_{10}$ is hyperbolic (all 
its proper regular subalgebras are either finite or affine), while 
$E_{11}$ is not (its regular subalgebra $E_{10}$ is neither finite 
nor affine). Consideration of these two algebras is chiefly motivated 
by recent attempts to find out more about the still unknown symmetries 
underlying M-Theory. Following the discovery of a Kac-Moody billiard
and arithmetical chaos in superstring cosmology in \cite{DH}, a novel 
one-dimensional $\sigma$-model based on the formal coset ${E_{10}}/K({E_{10}})$ 
was proposed in \cite{DHN}, with $K({E_{10}})$ the maximal compact subgroup 
of ${E_{10}}$, and a precise identification was made in a BKL-type 
expansion between the bosonic supergravity fields at a given space 
point and their first spatial gradients on the one hand, and the 
first three levels of ${E_{10}}$ on the other (the emergence of ${E_{10}}$ 
in the dimensional reduction of $D\!=\!11$ supergravity\cite{CJS} 
to one dimension had been conjectured already long ago \cite{Julia}). 
An alternative proposal was made in \cite{West,West1}, where 
the rank-11 algebra $E_{11}$ is conjectured to be the fundamental 
symmetry of M-Theory. It is likewise based on a non-linear realization 
of the relevant ``Kac-Moody group'', but differs from \cite{DHN} in 
that it puts more emphasis on spacetime covariance. The results of
the present paper may serve to discriminate between the two proposals,
as our tables show very explicitly where $E_{11}$ deviates from ${E_{10}}$
(namely by all those representations in Table 2, for which the last
label $m^{10}$ is different from zero).

As in \cite{DHN} (as well as in \cite{OPR,CJLP,KNS,West,West1}) our 
investigation is based on a decomposition of these algebras 
w.r.t. to their maximal $sl(n,\Rn)$ subalgebras, namely the two 
algebras $A_9 \equiv sl(10,\Rn)$ and $A_{10} \equiv sl(11,\Rn)$,
both of which can be enlarged to the Lie algebras of the full general 
linear groups $GL(10,\Rn)$  and $GL(11,\Rn)$ within ${E_{10}}$ and 
${E_{11}}$, respectively, by inclusion of the ``exceptional'' Cartan 
subalgebra generator. This decomposition is organized by means of 
the level $\ell$, which is the number of times a certain distinguished 
``exceptional'' root $\alpha_0$ appears in the decomposition of a 
given root $\alpha$ \cite{DHN}. The level-0 sectors then consist
of the $A_9$ and $A_{10}$ subalgebras themselves. For the first 
three levels, the representations which appear are of the same type 
for all $E_n$ with $n\geq 9$. The Dynkin labels of the relevant 
representations are 
\begin{eqnarray}\label{123}
\ell = 1 \;\; &:& \quad (001000000\dots) \nonumber\\
\ell = 2 \;\; &:& \quad (000001000\dots) \nonumber\\
\ell = 3 \;\; &:& \quad (100000010\dots) 
\end{eqnarray}
corresponding to the antisymmetric tensors with three and six indices
for $\ell=1,2$, respectively, and a mixed Young tableau for $\ell=3$;
therefore, the associated Lie algebra generators are 
$E^{a_1 a_2 a_3}, E^{a_1 \dots a_6}$ and $E^{a_0|a_1 \dots a_8}$
(with the constraint $E^{[a_0|a_1 \dots a_8]}=0$). In the physical 
interpretation, the first two of these are associated with the
electric and magnetic excitations of the 3-form field of maximal
$D\!=\!11$ supergravity, while the third can be viewed as dual 
to the graviton \cite{OPR,CJLP,Hull,West,DHN}. For the finite dimensional
Lie algebras $E_n$ with $n\leq 8$, there are no higher level 
representations beyond $\ell =3$; in particular, for $n=8$, the 
level-3 representation collapses to the $(1000000)$ of $A_7$, and for 
$E_7$ and $E_6$ it is absent altogether. For $n\geq 9$, there are
representations at all levels. While these are easy to classify for $E_9$,
a veritable explosion in the number of representations takes place
for $n\geq 10$, and this will be our main focus here. For the 
hyperbolic algebra ${E_{10}}$, we will obtain the complete list of 
$A_9$ representations up to level 18; for ${E_{11}}$, we give them up 
to level 10. In principle, these lists can be extended considerably 
further\footnote{Actually, the tables are now available up to 
level $\ell =28$ for ${E_{10}}$, where we have already 3276 inequivalent
irreducible representations of $SL(10,\Rn)$ with maximum outer
multiplicity equal to 46450629 for the representation $(110100011)$. 
Evident lack of space prevents us from presenting the complete
results here. However, all the raw data are included in the {\tt 
arXiv.org} source package of this work and can be found at
{\tt http://www.arxiv.org/e-print/hep-th/0301017}. This package
also contains the decomposition of the Feingold-Frenkel algebra
$AE_3$ in terms of $A_2$ representations up to level $\ell = 56$.
Code to calculate such decompositions has been made publicly available
\cite{Fischbacher:2002fr}.}.

There is by now a substantial body of work on root multiplicities
of low rank hyperbolic KM algebras, and the corresponding decompositions
of these algebras w.r.t. to certain distinguished affine subalgebras,
starting with the rank 3 hyperbolic algebra studied \cite{FF} where 
closed formulas were derived for multiplicities of roots of affine 
level $\leq 2$. This work was extended up to affine level 5 
and to other low rank algebras in \cite{BKM1}-\cite{BKM3} and 
\cite{Ka1}-\cite{Ka5}, \cite{KaM1,KaM2}. A closed formula for the 
${E_{10}}$ root multiplicities at affine level two was derived 
in \cite{KMW} (affine levels zero and one being trivial).
The decomposition w.r.t. an affine subalgebra has the advantage of
classifying an infinite number of Lie algebra elements, but only 
``close to the lightcone'', and appears rather hopeless beyond the
very first levels. By contrast, the decomposition w.r.t. to a finite 
dimensional algebra yields only a finite amount of information at 
each step, but can be carried to rather high levels. 
Besides, it is of special interest for possible applications 
in string and M-Theory because it yields a concrete realization of 
the algebra in terms of irreducible representations of subgroups 
that have a direct physical interpretation. To penetrate the structure 
of ${E_{10}}$ even further one can try to ``intertwine'' the representation 
theory of $A_9$ and $E_9$, or to exploit the decomposition under its 
principal $sl(2,\Rn)$ subalgebra \cite{Hu,NO}. As for ${E_{11}}$, and apart
from the $\ell\leq 3$ results of \cite{West,West1} (which do not require
knowledge of non-trivial ${E_{11}}$ root multiplicities), our results are
completely new; in particular, this is the first time that
non-trivial ${E_{11}}$ root multiplicities have been calculated.

While our methods are thus by no means sufficient to arrive at a completely 
explicit and manageable realization of these indefinite KM algebras, the 
results of this paper can nevertheless serve as useful ``raw data'' for 
future investigations. At the very least, they give a very concrete flavor 
of the stunning complexity of these Lie algebras: readers should keep 
in mind that the tables presented here represent only the very first 
steps into a mathematical structure, whose complexity will forever 
continue to grow with increasing level as one moves deeper 
and deeper into the lightcone. Even restricting attention to the 
low level representations exhibited here, the task of working out, 
say, the Lie algebra structure constants beyond the very first levels
seems too daunting even to contemplate. From the physics perspective,
the possible link noticed in \cite{DHN} between the three infinite 
towers of representations (\ref{n123}) and the spatial gradients
of the bosonic fields of $D\!=\!11$ supergravity is a tantalizing 
hint, but the task of finding a physical interpretation for all the 
other representations remains a major challenge. Realizing the full ${E_{10}}$
(or ${E_{11}}$) symmetry probably requires {\it a theory beyond} $D\!=\!11$ 
supergravity, as it seems very unlikely that these other representations 
simply correspond to auxiliary or St\"uckelberg-type degrees of freedom. 
For the time being, however, we must refer readers to the still unknown 
chapter in THE BOOK \cite{BOOK} for an answer to these and other questions!

\section{Roots and Dynkin labels}
\begin{figure}
\includegraphics{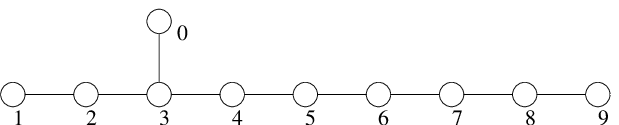}
\caption{\label{e10dynkin}$E_{10}$}
\end{figure}

\begin{figure}
\includegraphics{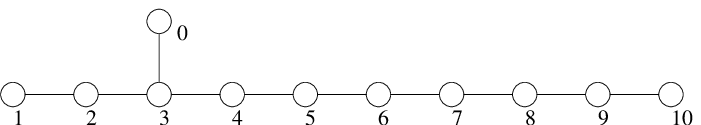}
\caption{\label{e11dynkin}$E_{11}$}
\end{figure}

The indefinite Kac Moody algebras $E_{10}$ and $E_{11}$ are 
characterized by the Dynkin diagrams displayed in figs.\ref{e10dynkin}
and  \ref{e11dynkin}. The simple roots of the respective $A_n$ 
subalgebras will be labeled $\alpha_1, \alpha_2,\dots$ (counting from 
left to right); the ``exceptional'' root connected to $\alpha_3$ will 
be designated by $\alpha_0$. Because both algebras are simply laced, 
the associated Cartan matrices are $A_{ij} = (\alpha_i | \alpha_j)$.

As in \cite{DHN} we will decompose both algebras into irreducible
representations of their finite dimensional subalgebras $A_9$ and 
$A_{10}$, respectively. This corresponds to a slicing of the 
forward lightcone in the respective root lattices of ${E_{10}}$ and ${E_{11}}$
by spacelike hyperplanes; hence the sections will be ellipsoidal
with {\em finitely many roots in each slice}. By contrast, slicing
the lightcone by lightlike or timelike hyperplanes would result
in a decomposition w.r.t. an affine or an indefinite subalgebra, 
in both of which cases each slice contains {\em infinitely many roots}. 
As already mentioned the corresponding $SL(n+1,\Rn)$ groups 
are enlarged to $GL(n+1,\Rn)$, respectively, by inclusion of the
Cartan generator $h_0$ associated with the exceptional root.

For both algebras any positive root can be expressed as
\begin{equation}\label{root}
\alpha = \ell\alpha_0 + \sum_{j=1}^n m^j \alpha_j  
\end{equation}
with non-negative integers $\ell$ and $m^j$, and $n=9$ or $n=10$
for ${E_{10}}$ and ${E_{11}}$, respectively. The number $\ell$ is the ``level''
and counts the number of occurrences of the exceptional root $\alpha_0$ 
in $\alpha$. It must not be confused with the ``affine level'' ($=m^9$) 
used in the decomposition of $E_{10}$ w.r.t. its affine subalgebra 
$E_9$ \cite{FF,KMW} (there appears to be no useful analog of the 
affine level for $E_{11}$). 

The irreducible representations (or ``modules'' in more
mathematical language) of $A_9$ and $A_{10}$ are completely 
characterized their Dynkin labels, which are 9-tuples 
$(p_1 \dots p_9)$ and 10-tuples $(p_1 \dots p_{10})$
of non-negative integers, respectively, such that the 
corresponding highest weights $\lambda$ are given by
\begin{equation}
\lambda = \sum_{j=1}^n p_j \lambda^j
\end{equation}
for $n=9,10$. Here $\lambda^j$ are the fundamental weights of $A_n$  
obeying $(\alpha_i | \lambda^j) = \delta_i^j$ 
with $i,j=1,\dots,n$ for $n=9,10$, respectively. At the same time, 
$p_j$ is equal to the number of columns with $j$ vertical boxes 
in the associated Young tableau. The fundamental weights of $A_n$ 
are explicitly given by $\lambda^i = \sum_{j=1}^n S^{ij} \alpha_j$, where 
$S^{ij}$ is the inverse Cartan matrix of $A_n$ (see below). We use small 
Greek letters for the weights of $A_n$ to distinguish them from the weights 
of $E_{n+1}$, whose fundamental weights $\Lambda^j$ (for $j=0,1,\dots ,n$)
must in addition satisfy $(\alpha_0 | \Lambda^i) = \delta_0^i$
and  differ by a vector orthogonal to the simple roots of $A_n$, 
see below. The explicit relation between the coefficients $m^j$ 
and the Dynkin labels will be derived below.

Each root $\alpha$ comes with a certain root multiplicity ${{\rm mult}} (\alpha)$
counting the number of independent Lie algebra elements associated
with $\alpha$ \cite{Kac}. These Lie algebra elements will be split up according
to which representation of $A_9$ or $A_{10}$ they belong to. Any root 
$\alpha$ will therefore also appear as a weight in certain representations 
${\mathcal R}$, with corresponding weight multiplicities ${{\rm mult}}_{\mathcal R} (\alpha)$. 
Furthermore, any such representation can appear several times, and 
the number of times a representation occurs within a given level 
will be called the {\em outer multiplicity}; the outer multiplicity 
of an admissible representation (i.e. a representation satisfying 
the two conditions (\ref{mi}) and (\ref{La2}) below) may be zero. 
We thus have the following picture: the finitely many roots at a given 
level $\ell$ are superpositions of certain $A_n$ weight diagrams, each 
one of which may appear several times with a certain outer multiplicity. 

While the relevant representations that can appear are not hard to 
find, the determination of the outer multiplicities is a difficult and 
as yet unsolved problem. Knowing them in closed form would be equivalent
to a complete knowledge of the root multiplicities for indefinite and 
hyperbolic Kac Moody algebras (there is so far not a single example of 
such an algebra for which the root multiplicities are known in closed form)
and would take us a considerable way towards the ultimate goal of
finding an explicit and manageable realization of the full algebra.
All we know is that typically the root multiplicities increase 
exponentially with their length $(-\alpha^2)$ \cite{Kac}, implying similar 
growth properties for the outer multiplicities.

In fact, the formulas restricting the possible representations
that can appear at level $\ell$ are easy to deduce \cite{DHN} and also 
easy to solve, especially with the help of a computer. To begin with, 
the representations which appear at level $\ell +1$ must be contained 
in the set of representations obtained by taking the product of the 
level one representation $(001000000)$ with all the representations
occurring at level $\ell$. This is because the level-$(\ell+1)$
generators are generally obtained by commuting the level-$\ell$ 
generators with a level one generator, and this is also how one
quickly derives the representations given in formula (\ref{123}).
However, many of the representations obtained in this way will drop
out, and the method becomes more and more impractical with higher levels. 

A better and much quicker way to derive the admissible representations
is by judicious analysis of the corresponding highest weights.
Indeed, the formula linking the  coefficients $m^j$ and the Dynkin 
labels of the relevant $A_9$ or $A_{10}$ is quite straightforward 
to deduce. First of all, we notice that the Chevalley generator $f_0$ 
is a highest weight state w.r.t. $A_9$ or $A_{10}$, with Dynkin label 
$(001000000\dots)$ because
\begin{eqnarray}
h_i (f_0) &\equiv& [h_i \, , f_0 ] = \delta_{i3} f_0 \nonumber\\
e_i (f_0) &\equiv& [e_i \, , f_0 ] = 0
\end{eqnarray}
for $i=1,\dots , n$. The associated highest weight is $\Lambda = - \alpha_0$.
Other highest weight states associated with roots 
$\Lambda\equiv - \alpha = -\alpha_{j_1} - \dots - \alpha_{j_k}$ are built from 
multiple commutators 
\begin{equation}
f_{j_1 \dots j_k} := [ f_{j_1} , \dots [f_{j_{k-1}},f_{j_k}]\dots ]
\end{equation}
by permuting the Chevalley generators $f_j$ in all possible ways
and taking suitable linear combinations
\begin{equation}
f^{(\Lambda)} = \sum c^{(\Lambda)}_{j_1 \dots j_k} f_{j_1 \dots j_k}
\end{equation}
which are annihilated by the adjoint action of all $e_i$ for
$i=1,\dots ,n$, i.e.
\begin{equation} 
e_i (f^{(\Lambda)}) \equiv [e_i , f^{(\Lambda)}] = 0
\end{equation}
The level of $f^{(\Lambda)}$ is the number of $j_\nu$ with $j_\nu = 0$.
The $A_n$ module is then built by acting on this state with the Chevalley
generators $f_i$, where again $i =1,\dots, n$. The weight of $f^{(\Lambda)}$ 
follows from
\begin{equation}\label{pi}
[h_i, f_{j_1 \dots j_k}] = \left( - \sum_{\nu=1}^k  A_{ij_\nu} \right) 
f_{j_1 \dots j_k}  = p_i f_{j_1 \dots j_k}
\end{equation}
An explicit basis of the module $V(\Lambda)$ is given in \cite{Verma}: all
states can be represented in the form
\begin{equation}
[f_{m_1},...,[f_{m_s}, f_{j_1 \dots j_k}]...] =
 f_{m_1 \dots m_s j_1 \dots j_k} \;\; , \quad m_\nu \in \{1,\dots ,n\}
\end{equation}
with weight labels
\begin{equation}
q_j = p_j - \sum_{\nu=1}^s A_{j m_\nu}
\end{equation}
where the $m_\nu$ are ordered according to the prescription given 
in \cite{Verma} (which is somewhat cumbersome to disentangle for
larger rank).

Setting 
$\alpha = \alpha_{j_1} + \dots + \alpha_{j_k} = \ell \alpha_0 + m^1 \alpha_1 + \dots + m^n \alpha_n$
and writing out the relation $p_i = - \sum_\nu A_{ij_\nu}$ we get 
(for $n=9,10$)
\begin{eqnarray}
p_1 &=& m_2 - 2m_1 \nonumber\\
p_2 &=& m_1 + m_3 - 2 m_2 \nonumber\\
p_3 &=& \ell + m_2 + m_4 - 2 m_3 \nonumber\\
p_4 &=& m_3 + m_5 - 2 m_4 \nonumber\\
&\cdot&\nonumber\\&\cdot&\nonumber\\&\cdot&\nonumber\\
p_{n-1} &=& m_{n-2} + m_n - 2 m_{n-1} \nonumber\\
p_n &=& m_{n-1} - 2 m_n
\end{eqnarray} 
Because this is a highest weight state by assumption, the corresponding
weight $\sum p_j \lambda^j$ is dominant, hence $p_j\geq 0$. Inverting this 
formula, we obtain \cite{DHN}
\begin{equation}\label{mi}
m^i = S^{i3} \ell - \sum_{j=1}^n S^{ij} p_j
\end{equation}
where $S^{ij}$ are the inverse Cartan matrices of $A_9$ and $A_{10}$,
respectively. This formula also allows us to express the highest
weight $\Lambda= -\alpha$ in terms of the $A_n$ weights: multiplying by
$\alpha_i$ and summing over $i$, we get
\begin{equation} 
\sum_{j=1}^n m^j \alpha_j = \ell \sum_{j=1}^n S^{3j} \alpha_j - 
                        \sum_{j=1}^n p_j \lambda^j
\end{equation}
where we have also used the definition of the fundamental $A_n$ weights.
Hence,
\begin{equation}
\Lambda \equiv - \alpha = -\ell \Big( \alpha_0 + \sum_{j=1}^n S^{3j} \alpha_j \Big) 
                  +  \sum_{j=1}^n p_j \lambda^j
\end{equation}
Observe that $p_j=(\alpha_j |\Lambda )= (\alpha_j |\lambda)$ for $j=1,\dots ,n$ because 
the first term has vanishing scalar product with these simple roots.

The inverse Cartan matrices are, respectively, given by
\begin{equation}
S^{ij}[A_9] = \frac1{10}
\left(\begin{array}{ccccccccc}
   9 & 8 & 7 & 6 & 5 & 4 & 3 & 2 & 1 \\
   8 & 16 & 14 & 12 & 10 & 8 & 6 & 4 & 2 \\
   7 & 14 & 21 & 18 & 15 & 12 & 9 & 6 & 3 \\
   6 & 12 & 18 & 24 & 20 & 16 & 12 & 8 & 4 \\
   5 & 10 & 15 & 20 & 25 & 20 & 15 & 10 & 5 \\
   4 & 8 & 12 & 16 & 20 & 24 & 18 & 12 & 6 \\
   3 & 6 & 9 & 12 & 15 & 18 & 21 & 14 & 7 \\
   2 & 4 & 6 & 8 & 10 & 12 & 14 & 16 & 8 \\
   1 & 2 & 3 & 4 & 5 & 6 & 7 & 8 & 9
\end{array}\right)
\end{equation}
for $A_9$ and
\begin{equation}
S^{ij} [A_{10}]= \frac1{11}
\left(\begin{array}{cccccccccc}
   10 & 9 & 8 & 7 & 6 & 5 & 4 & 3 & 2 & 1 \\
   9 & 18 & 16 & 14 & 12 & 10 & 8 & 6 & 4 & 2 \\
   8 & 16 & 24 & 21 & 18 & 15 & 12 & 9 & 6 & 3 \\
   7 & 14 & 21 & 28 & 24 & 20 & 16 & 12 & 8 & 4 \\
   6 & 12 & 18 & 24 & 30 & 25 & 20 & 15 & 10 & 5 \\
   5 & 10 & 15 & 20 & 25 & 30 & 24 & 18 & 12 & 6 \\
   4 & 8 & 12 & 16 & 20 & 24 & 28 & 21 & 14 & 7 \\
   3 & 6 & 9 & 12 & 15 & 18 & 21 & 24 & 16 & 8 \\
   2 & 4 & 6 & 8 & 10 & 12 & 14 & 16 & 18 & 9 \\
   1 & 2 & 3 & 4 & 5 & 6 & 7 & 8 & 9 & 10
\end{array}\right)
\end{equation}
for $A_{10}$. Because the entries of $S^{ij}$ are all positive, and 
because $m^i$ and $p_i$ must all be non-negative integers, the formula
(\ref{mi}) strongly constrains the possible representations for each $\ell$.

A further constraint derives from the fact that the highest weight
$\Lambda = -\alpha$ must be a root; this means that its norm must obey 
$\Lambda^2\leq 2$ \footnote{This is only a necessary but not a sufficient 
criterion for $\Lambda$ to be a root, see our discussion of the ${E_{11}}$
roots in the next section.}.This requirement effectively implements 
the Serre relations. For $E_{10}$ it implies
\begin{equation}\label{La2}
\Lambda^2 = \alpha^2 = \sum_{i,j=1}^9 p_i S^{ij} p_j - \ft1{10}\ell^2 \;\;
\Longrightarrow  \;\; \sum_{i,j=1}^9 p_i S^{ij} p_j \leq 2 + \ft{1}{10}\ell^2
\end{equation}
For $E_{11}$, we have the bound
\begin{equation}
\Lambda^2 = \alpha^2 = \sum_{i,j=1}^{10} p_i S^{ij} p_j - \ft2{11}\ell^2 \;\;
\Longrightarrow  \;\; \sum_{i,j=1}^{10} p_i S^{ij} p_j \leq 2+\ft{2}{11}\ell^2
\end{equation}
As we will see this constraint again eliminates many further
representations which would still be compatible with (\ref{mi}).

An important infinite series of ${E_{10}}$ Lie algebra elements was identified 
in \cite{DHN} by searching for ``affine representations'', namely 
those highest weights $\Lambda$ for which $m^9 =0$ in (\ref{root}).
There are three infinite towers of ${E_{10}}$ Lie algebra elements at levels 
$\ell=3k+1, 3k+2$ and $3k+3$ (for $k\geq 0$) with Dynkin labels 
\begin{eqnarray}\label{n123}
\ell = 3k+1 \;\; &:& \quad (00100000k) \nonumber\\
\ell = 3k+2 \;\; &:& \quad (00000100k) \nonumber\\
\ell = 3k+3 \;\; &:& \quad (10000001k) 
\end{eqnarray}
In our tables, these representations are always listed at the very top.
Inserting the Dynkin labels into the above formula we see that
the associated highest weights obey $\Lambda^2 =2$ for all $k$; since 
${\rm mult} (\alpha) = 1$ for $\alpha^2 =2$, the outer multiplicities
are always one. Explicitly, the Lie algebra elements are
\begin{equation}\label{affine1}
{E_{a_1 \dots a_k}}^{b_1 b_2  b_3} \;\; , \quad
{E_{a_1 \dots a_k}}^{b_1 \dots  b_6} \;\; , \quad
{E_{a_1 \dots a_k}}^{b_0 |b_1 \dots b_8}
\end{equation}
and were tentatively associated in \cite{DHN} to the spatial gradients
of the bosonic fields of $D=11$ supergravity and their duals. There exist
three similar infinite towers  of conjugate (or ``transposed'') Lie 
algebra elements obtained by acting with the Chevalley involution 
on the above elements. These affine representations are 
distinguished because they contain the affine subalgebra $E_9$. 
To see this more explicitly, we note that the elements of the latter 
must commute with the central charge of $E_9$. This requirement is 
equivalent to imposing a ``dimensional reduction'' on the elements
(\ref{affine1}) by setting $a_1 = \dots = a_k = 10$ and taking the 
remaining indices $b_j \in \{1,\dots , 9 \}$. Using the decomposition 
${\bf 248} \rightarrow {\bf 80} \oplus {\bf 84} \oplus \overline{\bf 84}$ 
of the adjoint of $E_8$ under its $SL(9,\Rn)$ subgroup, the relevant 
Lie algebra elements are then obtained from this truncated set and 
the generators of the subgroup $GL(9,\Rn)\subset GL(10,\Rn)$; the 
central charge generator $c$ is the remaining diagonal generator 
of $GL(10,\Rn)$.

A second infinite series of admissible affine $A_9$ representations 
at levels $\ell =3k$ is characterized by the Dynkin labels
\begin{equation}
\ell = 3k \;\; : \quad (00000000k)
\end{equation}
It is easy to see that these representations must have 
outer multiplicity zero. The associated roots are
\begin{equation}
\alpha \equiv k \Big( 3 \alpha_0 + \sum_{j=1}^9 m^j \alpha_j \Big) = k \delta
\end{equation}
where the root $\delta$ associated with the values 
$(m^1 \dots m^9) = (246543210)$ is the affine null vector of $E_9$.
However, there can be no such generator in $E_9$, because all the $E_9$ 
elements are already contained in the three infinite towers (\ref{n123})
and their transposed towers as we have just shown. 

At low levels the ``disappearance'' of certain representations can 
be understood as a consequence of the Jacobi identity, which implies 
that any commutator of the form
\begin{equation}\label{Jacobi}
\Big[ x_1, \dots , \big[ x_k , 
[E^{[a_1 a_2 a_3} ,[ E^{a_4 a_5 a_6}, E^{a_7 a_8 a_9]}]] \big] \dots \Big]
\end{equation}
must vanish for arbitrary Lie algebra elements $x_j$. For instance,
it is easy to see that the representation $(000000001)$ drops
out by virtue of (\ref{Jacobi}). At higher levels, unfortunately, the 
situation is rather less transparent, and it does not appear that
(\ref{Jacobi}) is of much use in eliminating representations, except
when it corresponds to a highest weight state for $A_n$.

For ${E_{11}}$, there are three similar infinite towers with labels
\begin{eqnarray}\label{nn123}
\ell = 3k+1 \;\; &:& \quad (00100000k0) \nonumber\\
\ell = 3k+2 \;\; &:& \quad (00000100k0) \nonumber\\
\ell = 3k+3 \;\; &:& \quad (10000001k0) 
\end{eqnarray}
and again $\Lambda^2 =2$ with outer multiplicity one for all these
representations. Now, the Lie algebra elements are given by
\begin{equation}
{E_{[a_1 b_1]\dots [a_k b_k]}}^{c_1 c_2  c_3} \;\; , \quad
{E_{[a_1 b_1]\dots [a_k b_k]}}^{c_1 \dots  c_6} \;\; , \quad
{E_{[a_1 b_1]\dots [a_k b_k]}}^{c_0 |c_1 \dots c_8}
\end{equation}
with antisymmetric index pairs $[a_1 b_1]$, etc. Obviously the 
representations (\ref{n123}) can again be obtained from those of 
(\ref{nn123}) by dimensional reduction (i.e. setting $c_i =11$
and taking $a_i, b_i \in \{1, \dots ,10\}$). However, a physical 
interpretation of these ${E_{11}}$ elements analogous to the one suggested 
in \cite{DHN} for the representations (\ref{n123}) appears no longer possible.

\section{Determination of low level $A_n$ representations}

We next explain how to work out the higher level representations 
that appear in the decompositions of ${E_{10}}$ and ${E_{11}}$ w.r.t. their respective 
$A_n$ subalgebras. Our results for the representations up to level $\ell =18$
for ${E_{10}}$ and up to level $\ell =10$ for ${E_{11}}$ are displayed in the tables 
following this section. The first step in determining 
the representations is to exploit the two conditions (\ref{mi}) 
and (\ref{La2}), which are already quite restrictive by themselves, 
because the $n$-tuples $(m^1...m^n)$ and $(p_1 \dots p_n)$ must both 
consist of non-negative integers. Hence (\ref{mi}) and (\ref{La2}) 
admit only a finite number of solutions for given $\ell$, both for 
${E_{10}}$ and ${E_{11}}$. Furthermore, with the help of a computer these solutions 
are easy to find up to very high levels.

For given $\ell$, each highest weight $\Lambda$ generates a representation
with a corresponding finite weight diagram 
$P(\Lambda)$ of $A_n$, which belongs to an elliptic 
slice of the forward lightcone in the root lattice of $E_{n+1}$.
The weights $\lambda\in P(\Lambda)$ come with certain weight multiplicities 
which we designate by ${{\rm mult}}_{\mathcal R} (\lambda)$ for the representation
${\mathcal R} \equiv {\mathcal R}(\Lambda)$. Hence the roots $\alpha$ at a given level make up 
a ``stack'' of weight diagrams, such that the root multiplicity 
${{\rm mult}} (\alpha)$ of $\alpha$ as a root of $E_{n+1}$ is the sum of the multiplicities 
of $\alpha$ as a weight occurring in the various weight diagrams $P(\Lambda)$. 
Thus the following formula linking the $E_{n+1}$ root multiplicities 
with the $A_n$ weight multiplicities is self-evident
\begin{equation}\label{Mult}
{{\rm mult}} (\alpha) = \sum_{i=1}^{n_\ell} \mu_\ell \Big({\mathcal R}_i^{(\ell)}\Big) \;
{{\rm mult}}_{{\mathcal R}_i^{(\ell)}} (\alpha)
\end{equation}
where ${\mathcal R}_i^{(\ell)}$ is the $i$-th representation at level $\ell$, and 
the index $i=1,\dots, n_\ell$ counts the level-$\ell$ representations.
The numbers $\mu_\ell ({\mathcal R}_i^{(\ell)})$ are the outer multiplicities 
at level $\ell$: they count how many times the representation 
${\mathcal R}_i^{(\ell)}$ occurs in the decomposition at level $\ell$.
As far as we know, there does not exist an analog of the Peterson 
recursion formula for the outer multiplicities, which is why we have to 
take the detour via (\ref{Mult}) for their determination.

The only quantities entering formula (\ref{Mult}) which are known 
in full generality, although somewhat cumbersome to work out for
more complicated representations, are the weight multiplicities  
${{\rm mult}}_{\mathcal R} (\lambda)$ for the $A_n$ representations ${\mathcal R}$. Given
an $A_n$ module ${\mathcal R} (\Lambda)$ with highest weight $\Lambda$ 
and any weight $\lambda\in P(\Lambda)$ (which implies 
$\lambda < \Lambda$) the weight multiplicity ${{\rm mult}}_{{\mathcal R}(\Lambda)}(\lambda)$
follows from the Freudenthal recursion formula (see for 
instance \cite{H})

\begin{eqnarray}\label{Freudenthal}
\Big( (\Lambda|\Lambda) + 2 \, {\rm ht}\, \Lambda - (\lambda|\lambda) - 
2 \, {\rm ht}\, \lambda \Big) \, {{\rm mult}}_{{\mathcal R}(\Lambda)}(\lambda)&&\nonumber\\
= 2 \sum_{\alpha >0} \sum_{k\geq 1} (\lambda + k\alpha | \alpha)\;{{\rm mult}}_{{\mathcal R}(\Lambda)}(\lambda + k\alpha)
\end{eqnarray}
where the first sum on the r.h.s. ranges over all positive roots of $A_n$.
For the calculation we employ a COMMON LISP implementation of an 
improved version of this formula, using the algorithm described in
\cite{MP82, Ste}. To determine the outer multiplicities of the 
relevant $A_n$ modules from the $E_{n+1}$ root  multiplicities (as 
far as they are known), we proceed by the method of exhaustion, 
starting with the highest weights obeying $\Lambda^2 =2$ whose 
associated representations always come with outer multiplicity 
one if $\Lambda$ is a real root (this is always true for ${E_{10}}$,
but there may be ``spurious real roots'' for ${E_{11}}$, see below).
Next, we descend the weight system until we hit a weight 
$\alpha\equiv\lambda\in P(\Lambda)$ whose multiplicity as a root of $E_{n+1}$ 
exceeds its multiplicity as a weight in $P(\Lambda)$. This root, then,
is a highest weight in a new representation. Consequently, any highest 
weight $\Lambda_1$ for which $\Lambda_1^2 < \Lambda_2^2$ may appear as a weight in 
the highest weight module of $\Lambda_2$. However, highest weights $\Lambda$ 
and $\Lambda'$ of the same length associated with different labels 
$(p_1 \dots  p_n)$ and  $(p'_1 \dots  p'_n)$ can never appear 
as weights in each other's weight diagram, as this would require 
$p_i\leq p'_i$ with strict inequality for at least one $i$, 
in contradiction with the assumed equality of the norms, implying 
$\sum p_i S^{ij} p_j = \sum p'_i S^{ij} p'_j$. 

The $E_{n+1}$ root multiplicities can be calculated in principle
from the Peterson recursion formula (see \cite{Kac}, Exercise 11.12). 
Unfortunately, this formula becomes more and more impractical 
with increasing height due to the large number
of root decompositions one has to take into account, see below.
For ${E_{10}}$, closed formulas exist up to affine level 2 \cite{KMW}
(following earlier results for the rank 3 case in \cite{FF}).
An implicit formula for the ${E_{10}}$ root multiplicities at affine 
level three was derived in \cite{BB}, but we have not been able 
to extract any numbers from it. All ${E_{10}}$ root multiplicities 
up to affine level 6 and height 231 have been tabulated 
in \cite{BGN}. For ${E_{11}}$, on the other hand, tables of root 
multiplicities have been lacking in the literature so far.

Because the ${E_{10}}$ root multiplicities listed in \cite{BGN} are a 
crucial ingredient in our calculation, we have performed an 
independent check based on a ``brute force'' evaluation of the 
Peterson recursion formula rather than the Weyl-orbit 
method employed there (whose description in the appendix 
of \cite{BGN} seems somewhat obscure in retrospect), confirming
and extending the tables of \cite{BGN}\footnote{Which are now
available up to height 320.}. The Peterson formula reads
\begin{equation}\label{Peterson1}
\Big( (\alpha|\alpha) - 2 \, {\rm ht} \, (\alpha) \Big) c_\alpha =
\sum_{\stackrel{\beta', \beta'' > 0}{\beta' + \beta'' = \alpha}}
(\beta' | \beta'') c_{\beta'} c_{\beta''}
\end{equation}
where
\begin{equation}\label{Peterson2}
c_\beta:= \sum_{k\geq 1} \frac1{k} {{\rm mult}} \left( \frac{\beta}{k} \right)
\end{equation}
and thus requires the decomposition of a given root $\Lambda$ 
as a sum of two terms, both of which are linear combinations of 
the simple roots with non-negative coefficients (but not necessarily 
roots themselves). For ${E_{10}}$ this procedure boils down to separating 
the possible $\beta$'s into two classes. The first consists 
of the positive real roots $\beta$ or their integer multiples up 
to a given height, which is half the height of the root $\Lambda$ 
whose multiplicity we wish to compute. A complete list of these 
can be generated from the simple roots by applying height increasing 
Weyl reflections (every positive real root can be reached by such 
reflections, cf. Proposition 5.1.e of \cite{Kac}), and then 
taking all their integer multiples. The second set consists of 
the positive imaginary roots up to the given height. For ${E_{10}}$, this
calculation is greatly simplified by noting that for imaginary roots
the minimal height
element in any Weyl orbit belongs to the fundamental Weyl chamber. 
So we start from the positive linear combinations of the fundamental 
weights (see \cite{KMW}) below the given height, and again apply 
height increasing Weyl reflections until we reach the given height.
It is then easy (for the computer, at least) to enumerate all the
possible combinations and to evaluate their contribution to the
r.h.s. of (\ref{Peterson1}). We note that the determination of 
the outer multiplicities via the method described above provides
a consistency check on the root multiplicities obtained in this way.
Namely, we have checked ``experimentally'' that in general one gets 
negative outer multiplicities if the wrong root multiplicities
are used in (\ref{Mult}).

For ${E_{11}}$ matters are more involved, first of all because so far 
no tables of ${E_{11}}$ root multiplicities have been available, and 
secondly because of some additional subtleties which have no
counterpart in ${E_{10}}$ \footnote{Properties of these doubly overextended
indefinite Kac-Moody algebras are discussed in \cite{GOW}, where they
are called ``very extended''.}. The latter are mainly due to the fact that
the fundamental weight dual to the doubly overextended root $\alpha_{10}$
is spacelike and is no longer a root because it has positive {\em and} 
negative fractional coefficients when expressed in terms of the 
simple roots, as well as a disconnected support on the Dynkin 
diagram; explicitly,
\begin{equation}\label{Lambda10}
\Lambda_{10} = \ft32 \alpha_0 + \alpha_1 + 2\alpha_2 + 3 \alpha_3 + \ft52 \alpha_4 +
    2 \alpha_5 + \ft32 \alpha_6 + \alpha_7 + \ft12 \alpha_8 - \ft12 \alpha_{10}
\end{equation}
The remaining fundamental weights are timelike, except for $\Lambda_9$
which is lightlike (and in fact the same as for ${E_{10}}$), and also have
fractional coefficients which are, however, all positive with connected
support on the Dynkin diagram. A new phenomenon is the appearance of 
spurious ``real roots'': for instance, by acting on $2\Lambda_{10}$ with 
suitable Weyl reflections we can easily produce combinations which 
look like real roots (having positive integer coefficients and connected 
support on the Dynkin diagram), but in fact are not! In listing the 
solutions to (\ref{mi}) and (\ref{La2}) we must watch out for such 
spurious ``real roots'': the corresponding solutions in the table appear 
with vanishing ${E_{11}}$ root multiplicity, and therefore also vanishing 
outer multiplicity (again, negative outer multiplicities would result
if this subtlety were not taken into account). To be on the safe side 
we have applied brute force once more in evaluating (\ref{Peterson1})
by making lists of decompositions up to the full height of the root 
(and not just half the height as for ${E_{10}}$), whose multiplicity 
we are interested in. The relevant ${E_{11}}$ root multiplicities 
are listed in the penultimate column of Table 2. 
Of course, the ${E_{11}}$ root multiplicities for the ``hyperbolic roots'' 
(obeying $m^{10}=0$) coincide with their multiplicities as ${E_{10}}$ roots. 
In contradistinction to ${E_{10}}$, the root multiplicities start 
depending on the direction of $\alpha$ already for $\alpha^2 = -2$, 
where we have two Weyl orbits, with multiplicities 44 and 46, 
respectively (for higher rank algebras with more than 
one lightlike fundamental weight, this may already happen for 
null roots). Another noteworthy feature is that the ${E_{11}}$
root multiplicities tend to be less than the number of transversal 
polarizations, in contrast to ${E_{10}}$ where they are equal or greater. 
For instance, the null roots of ${E_{11}}$, all of which belong to a 
single Weyl orbit (and its integer multiples) have multiplicity 
equal to 8 instead of the number of transversal polarizations of 
a photon in eleven dimensions (=9), that one would expect on the basis 
of a vertex operator construction of such algebras \cite{GN}.

For the outer multiplicities, we also note some general features. First of 
all, the outer multiplicities of the $A_n$ representations in $E_{n+1}$ at 
level $\ell + 1$ usually will be strictly less than the outer multiplicities 
appearing in the product of the representations at level $\ell$ 
with the level one representation $(001000000)$. Secondly, there 
are very few representations that have vanishing outer multiplicities; 
e.g. for $E_{10}$, there are none at levels $\ell =7,11,13,14$ 
and 17. Thirdly, up to the maximum level we have checked, all the $A_9$ 
representations in ${E_{10}}$ with vanishing outer multiplicity have 
highest weights $\Lambda$ obeying $\Lambda^2 = 0$. Apart from the spurious
``real roots'' this is also true for $E_{11}$. In other words, 
if a representation can appear, it will appear, usually 
with an outer multiplicity that increases with the level $\ell$.

In the first set of tables we list the $A_9$ representations appearing
in the decomposition of ${E_{10}}$ up to and including level $\ell =18$
in lexicographic order $\{ m^j\}$ (starting from the back).
The level is indicated in the first column, and in the second and 
third columns we give the Dynkin labels and the coefficients 
$m^j$ appearing in (\ref{root}). The norms  $\Lambda^2$ of the associated 
highest weights are given in the fourth column. In the fifth 
column we list the dimension of the $A_9$ representation 
${\mathcal R}(\Lambda)\equiv{\mathcal R}_{(p_1 \dots p_9)}$, which are 
computed by means of the Weyl dimension formula 
\begin{equation}
{\rm dim}\, {\mathcal R}(\Lambda) = 
\prod_{\alpha > 0} \frac{(\Lambda + \rho | \alpha )}{(\rho | \alpha)}
\end{equation}
where $\rho$ is the $A_9$ Weyl vector such that the product
$(\rho|\alpha)$ is just the height $\sum_j m^j$ of $\alpha$ as a
root of $A_9$. The next column gives the multiplicity ${{\rm mult}} (\alpha)$ of 
the highest weight $\Lambda\equiv -\alpha$ as a root of ${E_{10}}$, taken 
from \cite{KMW} and \cite{BGN} (and checked again by our independent
calculation). The last column contains our main result, namely 
the outer multiplicities of the associated $A_9$ representations 
in ${E_{10}}$. The ${E_{11}}$ tables are organized similarly.

\section{Acknowledgments}

We are grateful to O.~B\"arwald for discussions and correspondence,
and to P.~Ramond for alerting us to refs. \cite{Verma,MP82}.
H.N. would like to thank the organizers of this conference, and 
N. Sthanumoorthy in particular, as well as the Center of Mathematical
Sciences in Chennai and T. Govindarajan for hospitality.

\newpage

{\small
\setlongtables

\begin{longtable}{|r|r|r|r|r|r|r|}
\caption{\label{e10table}$A_9$ representations in $E_{10}$ up to level $\ell=18$}\\
\hline
\multicolumn{1}{|c|}{$\ell$}&\multicolumn{1}{c|}{$p$}&\multicolumn{1}{c|}{$m$}&\multicolumn{1}{c|}{$\Lambda^2$}&\multicolumn{1}{c|}{dim ${\mathcal{R}}(\Lambda)$}&\multicolumn{1}{c|}{${\rm mult}(\Lambda)$}&\multicolumn{1}{c|}{$\mu$}\\
\hline
\hline
$1$&(\hbox to 0.5em{\hfill 0}\hbox to 0.5em{\hfill 0}\hbox to 0.5em{\hfill 1}\hbox to 0.5em{\hfill 0}\hbox to 0.5em{\hfill 0}\hbox to 0.5em{\hfill 0}\hbox to 0.5em{\hfill 0}\hbox to 0.5em{\hfill 0}\hbox to 0.5em{\hfill 0})&\kern-0.7em\hbox to 1.5em{\hfill 0}\hbox to 1.5em{\hfill 0}\hbox to 1.5em{\hfill 0}\hbox to 1.5em{\hfill 0}\hbox to 1.5em{\hfill 0}\hbox to 1.5em{\hfill 0}\hbox to 1.5em{\hfill 0}\hbox to 1.5em{\hfill 0}\hbox to 1.5em{\hfill 0}&2&120&1&1\\
\hline
$2$&(\hbox to 0.5em{\hfill 0}\hbox to 0.5em{\hfill 0}\hbox to 0.5em{\hfill 0}\hbox to 0.5em{\hfill 0}\hbox to 0.5em{\hfill 0}\hbox to 0.5em{\hfill 1}\hbox to 0.5em{\hfill 0}\hbox to 0.5em{\hfill 0}\hbox to 0.5em{\hfill 0})&\kern-0.7em\hbox to 1.5em{\hfill 1}\hbox to 1.5em{\hfill 2}\hbox to 1.5em{\hfill 3}\hbox to 1.5em{\hfill 2}\hbox to 1.5em{\hfill 1}\hbox to 1.5em{\hfill 0}\hbox to 1.5em{\hfill 0}\hbox to 1.5em{\hfill 0}\hbox to 1.5em{\hfill 0}&2&210&1&1\\
\hline
$3$&(\hbox to 0.5em{\hfill 1}\hbox to 0.5em{\hfill 0}\hbox to 0.5em{\hfill 0}\hbox to 0.5em{\hfill 0}\hbox to 0.5em{\hfill 0}\hbox to 0.5em{\hfill 0}\hbox to 0.5em{\hfill 0}\hbox to 0.5em{\hfill 1}\hbox to 0.5em{\hfill 0})&\kern-0.7em\hbox to 1.5em{\hfill 1}\hbox to 1.5em{\hfill 3}\hbox to 1.5em{\hfill 5}\hbox to 1.5em{\hfill 4}\hbox to 1.5em{\hfill 3}\hbox to 1.5em{\hfill 2}\hbox to 1.5em{\hfill 1}\hbox to 1.5em{\hfill 0}\hbox to 1.5em{\hfill 0}&2&440&1&1\\
\hline
\end{longtable}
}

\newpage

{\small
\setlongtables

\begin{longtable}{|r|r|r|r|r|r|r|}
\caption{\label{e11table}$A_{10}$ representations in $E_{11}$ up to level $\ell=10$}\\
\hline
\multicolumn{1}{|c|}{$\ell$}&\multicolumn{1}{c|}{$p$}&\multicolumn{1}{c|}{$m$}&\multicolumn{1}{c|}{$\Lambda^2$}&\multicolumn{1}{c|}{dim ${\mathcal{R}}(\Lambda)$}&\multicolumn{1}{c|}{${\rm mult}(\Lambda)$}&\multicolumn{1}{c|}{$\mu$}\\
\hline
\hline
$1$&(\hbox to 0.5em{\hfill 0}\hbox to 0.5em{\hfill 0}\hbox to 0.5em{\hfill 1}\hbox to 0.5em{\hfill 0}\hbox to 0.5em{\hfill 0}\hbox to 0.5em{\hfill 0}\hbox to 0.5em{\hfill 0}\hbox to 0.5em{\hfill 0}\hbox to 0.5em{\hfill 0}\hbox to 0.5em{\hfill 0})&\kern-0.7em\hbox to 1.5em{\hfill 0}\hbox to 1.5em{\hfill 0}\hbox to 1.5em{\hfill 0}\hbox to 1.5em{\hfill 0}\hbox to 1.5em{\hfill 0}\hbox to 1.5em{\hfill 0}\hbox to 1.5em{\hfill 0}\hbox to 1.5em{\hfill 0}\hbox to 1.5em{\hfill 0}\hbox to 1.5em{\hfill 0}&2&165&1&1\\
\hline
$2$&(\hbox to 0.5em{\hfill 0}\hbox to 0.5em{\hfill 0}\hbox to 0.5em{\hfill 0}\hbox to 0.5em{\hfill 0}\hbox to 0.5em{\hfill 0}\hbox to 0.5em{\hfill 1}\hbox to 0.5em{\hfill 0}\hbox to 0.5em{\hfill 0}\hbox to 0.5em{\hfill 0}\hbox to 0.5em{\hfill 0})&\kern-0.7em\hbox to 1.5em{\hfill 1}\hbox to 1.5em{\hfill 2}\hbox to 1.5em{\hfill 3}\hbox to 1.5em{\hfill 2}\hbox to 1.5em{\hfill 1}\hbox to 1.5em{\hfill 0}\hbox to 1.5em{\hfill 0}\hbox to 1.5em{\hfill 0}\hbox to 1.5em{\hfill 0}\hbox to 1.5em{\hfill 0}&2&462&1&1\\
\hline
$3$&(\hbox to 0.5em{\hfill 1}\hbox to 0.5em{\hfill 0}\hbox to 0.5em{\hfill 0}\hbox to 0.5em{\hfill 0}\hbox to 0.5em{\hfill 0}\hbox to 0.5em{\hfill 0}\hbox to 0.5em{\hfill 0}\hbox to 0.5em{\hfill 1}\hbox to 0.5em{\hfill 0}\hbox to 0.5em{\hfill 0})&\kern-0.7em\hbox to 1.5em{\hfill 1}\hbox to 1.5em{\hfill 3}\hbox to 1.5em{\hfill 5}\hbox to 1.5em{\hfill 4}\hbox to 1.5em{\hfill 3}\hbox to 1.5em{\hfill 2}\hbox to 1.5em{\hfill 1}\hbox to 1.5em{\hfill 0}\hbox to 1.5em{\hfill 0}\hbox to 1.5em{\hfill 0}&2&1760&1&1\\
\hline
\end{longtable}
}

\end{document}